\documentclass[preprint,prl,showpacs,preprintnumbers,amsmath,amssymb,superscriptaddress]{revtex4-1}

\usepackage{times}
\usepackage{graphicx}% Include figure files
\usepackage{dcolumn}% Align table columns on decimal point
\usepackage{bm}% bold math
\usepackage{chapterbib}
\usepackage{color}

\begin{document}

\preprint{}

\title{Coupled skyrmion sublattices in Cu$_{2}$OSeO$_{3}$}% Force line breaks with \\

\author{M.C. Langner}
\affiliation{Materials Science Division, Lawrence Berkeley National Laboratory, Berkeley, CA 94720, USA}
\author{S. Roy}
\thanks{Corresponding author \\
sroy@lbl.gov}
\affiliation{Advanced Light Source, Lawrence Berkeley National Laboratory, Berkeley CA 94720, USA}
\author{S. K. Mishra}
\affiliation{Advanced Light Source, Lawrence Berkeley National Laboratory, Berkeley CA 94720, USA}
\author{J. C. T. Lee}
\affiliation{Advanced Light Source, Lawrence Berkeley National Laboratory, Berkeley CA 94720, USA}
\author{X. W. Shi}
\affiliation{Advanced Light Source, Lawrence Berkeley National Laboratory, Berkeley CA 94720, USA}
\author{M. A. Hossain}
\affiliation{Advanced Light Source, Lawrence Berkeley National Laboratory, Berkeley CA 94720, USA}
\author{Y.-D. Chuang}
\affiliation{Advanced Light Source, Lawrence Berkeley National Laboratory, Berkeley CA 94720, USA}
\author{S. Seki}
\affiliation{RIKEN, Center for Emergent Matter Science, Wako 351-0198, Japan.}
\affiliation{PRESTO, Japan Science and Technology Agency, Tokyo 102-0075, Japan.}
\author{Y. Tokura}
\affiliation{RIKEN, Center for Emergent Matter Science, Wako 351-0198, Japan.}
\affiliation{Department of Applied Physics and Quantum Phase Electronics Center, University of Tokyo, Tokyo 113-8656, Japan.}
\author{S. D. Kevan}
\affiliation{Advanced Light Source, Lawrence Berkeley National Laboratory, Berkeley CA 94720, USA}
\affiliation{Department of Physics, University of Oregon, Eugene, OR 97401, USA.}
\author{R. W. Schoenlein}
\affiliation{Materials Science Division, Lawrence Berkeley National Laboratory, Berkeley, CA 94720, USA}

\date{\today}% It is always \today, today,

\begin{abstract}
We report the observation of a skyrmion lattice in the chiral multiferroic insulator Cu$_{2}$OSeO$_{3}$ using Cu L$_{3}$-edge resonant soft x-ray diffraction. We observe the unexpected existence of two distinct skyrmion sub-lattices that arise from inequivalent Cu sites with chemically identical coordination numbers but different magnetically active orbitals .  The skyrmion sublattices are rotated with respect to each other implying a long wavelength modulation of the lattice. The modulation vector is controlled with an applied magnetic field, associating this Moir\'{e}-like phase with a continuous phase transition. Our findings will open a new class of science involving manipulation of quantum topological states. \end{abstract}


\maketitle

\chapter{}
% ****** Start of file apssamp.tex ******
%
%   This file is part of the APS files in the REVTeX 4 distribution.
%   Version 4.0 of REVTeX, August 2001
%
%   Copyright (c) 2001 The American Physical Society.
%
%   See the REVTeX 4 README file for restrictions and more information.
%
% TeX'ing this file requires that you have AMS-LaTeX 2.0 installed
% as well as the rest of the prerequisites for REVTeX 4.0
%
% See the REVTeX 4 README file
% It also requires running BibTeX. The commands are as follows:
%
%  1)  latex apssamp.tex
%  2)  bibtex apssamp
%  3)  latex apssamp.tex
%  4)  latex apssamp.tex
%
%\documentclass[twocolumn,showpacs,preprintnumbers,amsmath,amssymb,superscriptaddress]{revtex4}
%\documentclass[twocolumn,prl,showpacs,amsmath,amssymb,superscriptaddress]{revtex4-1}

Topologically protected novel phases in condensed matter systems are a current research topic of tremendous interest due to both the unique physics and their potential in device applications. Skyrmions are a topological phase that in magnetic systems manifest as a hexagonal lattice of spin-vortices. Like its charge counterpart, the Majorana fermions in topological insulators \cite{top_review}, or superconductor junctions for quantum computing applications \cite{junc_review}, skyrmions are a potential candidate for spintronic-based information processing with the advantage that they can be moved coherently over macroscopic distances with very low currents \cite{Jonietz, Schulz}. Understanding the fundamental physics and mechanisms for controlling their dynamical properties are important scientific challenges. 

Skyrmions were first discovered in MnSi alloy using neutron scattering, which was followed by real space maps of these particle-like states using Lorentz TEM \cite{Muhlbauer, Yu}. A major breakthrough came when skyrmions were discovered in Cu$_{2}$OSeO$_{3}$ which is a multiferroic insulator \cite{Seki}. This opened up the possibility of manipulating the skyrmions with electric field, but at the same time, it also became imperative to understand how different orbitals stabilize the skyrmionic phase. In this report we exploit the orbital sensitive aspect of resonant x-rays to unravel the electronic origin of the skyrmions. As a function of incident photon energy a rotational splitting of the skyrmion satellite peaks was observed that could be controlled by an external magnetic field. The coupled response of these sub-lattices to external magnetic field suggests a secondary interaction term that has not been predicted.

Cu$_{2}$OSeO$_{3}$ is a cubic insulator with the same B20 space group symmetry as MnSi alloys. The local magnetic structure is primarily ferrimagnetic. The two ferrimagnetic sublattices are formed by spins on Cu-atoms with different oxygen bonding geometries: specifically Cu sites that are connected to square pyramidal or trigonal bipyramidal oxygen ligands \cite{Seki, Bos}. Helical magnetic phases arise from a combination of symmetric spin-exchange interactions, such as those that lead to ferromagnetism, and antisymmetric exchange resulting from a Dzyaloshinskii-Moriya interaction \cite{Binz, Yang}. In materials within a certain symmetry group - noncentrosymmetric, nonpolar, cubic materials - spiral spin phases develop along the three primary axes as a result of these interactions. Application of a magnetic field breaks the symmetry of the ground state, and results in a stable topological phase (within a narrow range of temperatures and applied fields) consisting of periodic magnetic vortices, known as the skyrmion phase. 

The insulating nature of Cu$_{2}$OSeO$_{3}$ makes it somewhat unusual in the family of magnetic skyrmion materials, in that the lack of conductivity means the mechanisms for controlling the skyrmions may differ from those in metallic skyrmion systems. Recent experiments have shown that in a metallic compound such as MnSi, the skyrmions can be manipulated using small electric currents \cite{Jonietz}.  Electric-field induced rotation of the skyrmion lattice has been previously demonstrated in Cu$_{2}$OSeO$_{3}$ \cite{White}, while measurements of the dielectric function indicate that the sample develops a spontaneous ferroelectric polarization along the [111] direction in response to an applied magnetic field in the same direction, which indicates some degree of magneto-electric coupling \cite{Seki, Bos}. Additionally, like MnSi, the skyrmion phase in Cu$_{2}$OSeO$_{3}$ can be stabilized only in a small region below the critical temperature illustrating the importance of strong spin-fluctuations in the creation and stabilization of the phase \cite{Muhlbauer, Seki}. 

Resonant x-ray scattering experiments were performed at the Advanced Light Source, Lawrence Berkeley National Laboratory. The scattering geometry and relative orientation of the applied magnetic field is shown in Fig 1(a). We tuned the incident sigma polarized x-ray beam to the Cu L3 edge and aligned the (001) lattice Bragg peak scattering signal into the detector. Figure 1(b) shows camera images of the diffraction peaks from the skyrmion phase at T = 57.5 K with a field of 20 mT applied in the vertical direction. The vertical field direction corresponds approximately to the (011) lattice direction in the diffraction geometry. The magnetic peaks appear as satellites at (0 0 1) $\pm \tau$ around the (0 0 1) lattice Bragg peak, where $\tau$ represents the q-vectors for the magnetic ordering. A beam block was placed in front of the camera to eliminate the (001) lattice peak, and an image taken above T$_{C}$ was subtracted to further enhance the contrast of the magnetic peaks.  We observed two fold symmetric satellite peaks around (0 0 1) $\pm \tau$, with $\tau =$ (0.103, 0, -0.04) nm$^{-1}$, arising due to the helical phase at T = 40 K and zero applied field.  

At T = 57.5 K and an applied magnetic field of 20 mT we observed the existence of the six-fold symmetric Bragg peak due to the skyrmion lattice (Fig 1(b)). The skyrmion structure forms in a plane perpendicular to the applied field direction \cite{Muhlbauer, Binz} and the diffraction peaks show a distortion along the sample surface due to the projection of the skyrmions onto the (001) plane. The size of the distortion is determined by the angle subtended between the surface normal and magnetic field direction as shown in Fig 1(a). This geometrical effect causes a reduction of the surface q-vector perpendicular to the x-ray direction, and leads to a compression of the six-fold symmetry in the camera image. Accounting for this distortion, the calculated q-vectors for both the helical and skyrmion phase correspond to a periodicity of 59.5 $\pm$ 5 nm, consistent with published values \cite{Schulz}. 

\begin{figure}
\label{fig:fig1}
\includegraphics[width=3.5in]{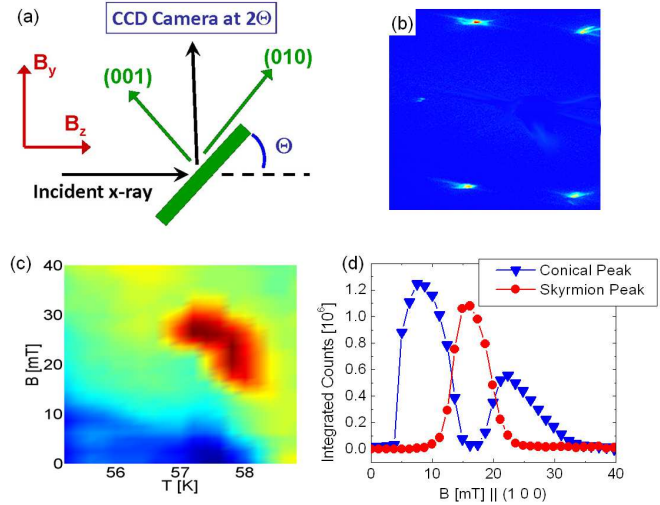}
\caption{(a) Schematic diagram of scattering geometry. The diffracted beam is captured by the CCD detector.  The magnetic field direction is fixed in the laboratory reference frame. (b) Camera image showing five skyrmion diffraction peaks at 57.5 K with a 20 mT field applied vertically (B${_y}$); the sixth peak is off camera. For clarity, the (001) lattice peak has been subtracted out.  (c) Phase diagram of the skyrmion (red) and helical (blue) peaks as a function of temperature and vertical applied field.  (d) Intensity of conical and skyrmion peaks as a function of field along (1 0 0) at 57.5 K. }
\end{figure}

Figure 1 (c) shows the helical/skyrmion phase diagram as determined by our scattering measurements with a vertical applied field. Figure 1 (d) shows the intensities of the helical, conical and skyrmion peaks for a field applied along the (100) lattice direction, with the sample initially in the helical state aligned along (001). Above the 5 mT de-pinning threshold, the conical phase aligns along (100), resulting in a peak at ($|\tau|$ 0 1).  At higher fields, the skyrmion phase appears as a peak at (0 0 1 + $|\tau|$) that results from scattering off of the skyrmion columns aligned along (100). Further increase in the applied field re-establishes the conical phase. We observed peaks from the skyrmion phase for all available orientations of the applied field, i.e. for field applied either along the x, y or z direction of the laboratory coordinate axes, thereby indicating the robustness of the phase.  

As a function of applied magnetic field, we observed that the peaks from the skyrmion structure split into two distinct sets, distinguished by the resonant photon energy, with both displaying six-fold symmetry. Figure 2(a) shows the full six-fold structure at a photon energy of 933 eV, where both sets of peaks are visible. Individual Bragg peaks within each set are separated by 60 degrees in the plane perpendicular to the applied field and remains consistent with the six-fold symmetry discussed above. The evolution of the double-peak structure is evident from the energy dependence of the two peaks, shown in figure 2(b-f). The maximum amplitudes of the peaks are separated in photon energy by ~ 2 eV which suggests that the two sets of peaks arise from the two inequivalent Cu (Cu$^{I}$ and Cu$^{II}$) sites in the Cu$_{2}$OSeO$_{3}$ unit cell. In addition to differences in Cu-Oxygen ligand geometry that affect the crystal field energy of the 3d orbitals, the inequivalent Cu sites have different magnetically active d-level orbitals; for the Cu$^{I}$ site the valence hole is d$_{z^{2}}$, while for Cu$^{II}$ the hole is d$_{x^{2}-y^{2}}$ \cite{Yang}. The probability matrix elements of the 2p $\to$ 3d transition are dependent on the spatial overlap of the core level and valence orbital. For the L$_{3}$ transition, the core levels are 2p$^{3/2}$ states that are further split (m$_{J}$ orbitals) by exchange and spin orbit interactions \cite{Rossi, Thole}. The difference in d-orbital symmetry between the Cu$^{I}$ and Cu$^{II}$ sites gives rise to dominant transitions from different m$_{J}$ states, resulting in a shift of the peak transition energy. These two sets of peaks are therefore representative of the two magnetic sites that make up the ferrimagnetic spin alignment in Cu$_{2}$OSeO$_{3}$. 

\begin{figure*}
\label{fig:fig2}
\includegraphics[width=6in]{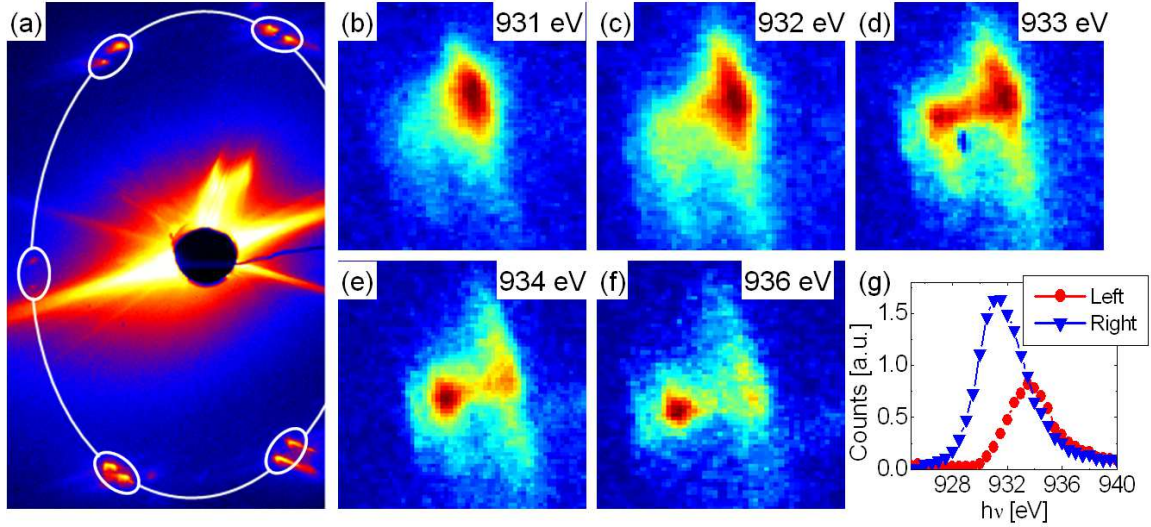}
\caption{Camera images showing dual-peak skyrmion structure. (a) Five sets of peaks, taken at 933 eV, are highlighted by the white ovals. The dotted line represents fixed $|\tau|$ contour. All the five pairs reside on the dotted line indicating that there exists in the sample two hexagonal lattices of skyrmions, each being offset by an in-plane rotation. (b)-(f) Evolution of the double peak as a function of the incident photon energy and fixed q. (h) Integrated peak intensities as a function of photon energy showing two different peak energies for the two rotated skyrmion lattices at fixed q.}
\end{figure*}

Along with the peak symmetry, figure 2(a) shows an arc representing constant amplitude of the skyrmion wavevector. All of the observed peaks fall on this arc, indicating that the periodicity of the skyrmion from each of the Cu sublattices is the same and that the symmetries of both peak sets are coplanar. The two sets of peaks therefore occur because one skyrmion lattice is rotated relative to the other in the plane perpendicular to the applied field.  This is consistent with the hierarchy of free-energy contributions observed in other skyrmion systems, where the $\langle M^{2} \rangle^{2}$ term in the free energy expansion favors a six-fold symmetry in the plane, but pinning of the in-plane angle of the six-fold structure is weak \cite{Muhlbauer}.

\begin{figure}
\label{fig:fig3}
\includegraphics[width=2.75 in]{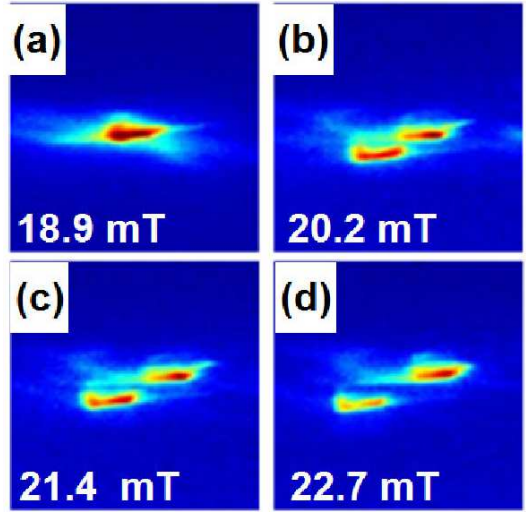}
\caption{Skyrmion peak splitting as a function of applied field. The data shown were taken at an incident x-ray energy of 932 eV.}
\end{figure}

\begin{figure}
\label{fig:fig4}
\includegraphics[width=3.5in]{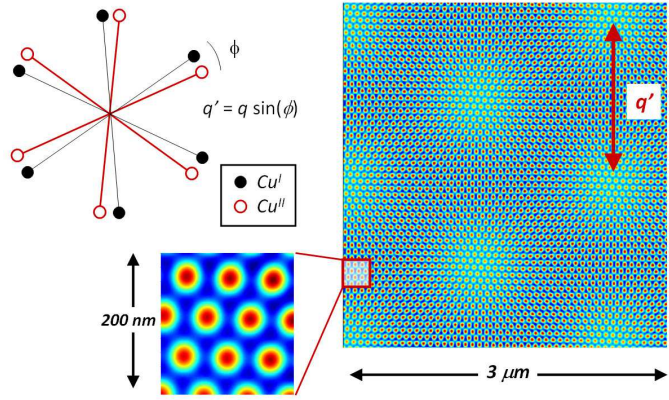}
\caption{Top left: Schematic of the relative rotated positions of Cu$^{I}$ and Cu$^{II}$ sublattice contributions to the x-ray diffraction images shown in figure 2.  Bottom inset: Spin configuration on the length scale of individual skyrmions (red up, blue down).  The local magnetization on this length scale is relatively constant, with only long-range changes to the ferrimagnetism.  Right: Calculated long-range spatial modulation of the overall spin along the magnetic field, corresponding to a Moire pattern arising from the relative rotation of two skyrmion lattices as shown in the diagram on the left.  The plot shows the spatial configuration corresponding to a 2.4 degree relative rotation. }
\end{figure}

In figure 3 we show the evolution of the skyrmion peak structure with applied magnetic field. At lower fields, a single peak is present; this peak separates into two at higher fields, with increasing separation as the field is increased. This is remarkable because now we can use a magnetic field as a tool to control the relative spatial configuration of the skyrmion sublattices. The rotation and superposition of two periodic structures creates a Moir\'{e} pattern, and therefore the in-plane rotation of the two skyrmion sublattices implies a long-wavelength variation in the local magnetic moment.
 
For small angular separation $\phi$ between anti-aligned magnetic sublattices $\vec{M_{1}}$ and $\vec{M_{2}}$, the magnetization along the applied field \textbf{B} can be written in cylindrical coordinates as:

\begin{widetext}
\[
\begin{aligned}
M_{z} (r, \theta) = \sum\limits_{n=1}^3 & [(M_{1} - M_{2})Cos(q r Cos(\frac{2 \pi n}{3} - \theta)) \\ &- (M_{1} + M_{2})Sin(q r Cos(\frac{2 \pi n}{3} - \theta)) Sin( q r Sin (\phi) Sin(\frac{2 \pi n}{3} - \theta))]
\end{aligned}
\]
\end{widetext}
 
Here \textbf{q} is the magnitude of the skyrmion wave vector, and \textbf{M$_{1}$} and \textbf{M$_{2}$} are the two ferrimagetic sublattice magnetizations, with magnitudes that occur in a three-to-one ratio in Cu$_{2}$OSeO$_{3}$. The sum represents the three q-vector directions. Note that the sublattice rotation preserves the six-fold symmetry, and the standard skyrmion lattice is recovered for \textbf{$\phi \to 0$}. The long-wavelength modulation of the magnetization is represented by the second term, and has a wavevector of \textbf{$q' = q Sin(\phi)$}. For the \textbf{$\phi$} measured in figure 3(d), this corresponds to a period of 1.2 microns, or approximately 20 periods of the skyrmion lattice. The resulting long-range structure is illustrated in figure 4. 

We propose the following explanation for the relative rotation of the skyrmion sublattices.  The Cu$_{2}$OSeO$_{3}$ unit cell contains 9 Cu$^{I}$ and 3 Cu$^{II}$ atoms with complex magnetic interactions \cite{Yang, Chizhikov}.  When a magnetic field is applied, the spins within the unit cell can lower the magnetic energy by canting or reorienting the spins along the applied field direction, with some cost to the local exchange energies.  We note that the exchange is frustrated, with an observed metamagnetic transition at fields similar to those at which we see the rotation of the skyrmion diffraction peaks \cite{Bos}.

As we observed the splitting changing as a function of magnetic field, we propose that canting of the magnetization or spin reorientation within the unit cell leads to a favorable long-range modulation of the ferrimagnetism resulting from spin-orbit interactions.  This implies a term with symmetry $M \times \delta M_{z}\left( B \right)$ in the local energy that tilts the spin plane of one sublattice relative to the other (see supplemental for further discussion) \cite{Supplemental}.  Additionally, higher-order terms in the phenomenological Ginzburg-Landau expansion related to the spin-orbit coupling favor modulations of the ferrimagnetism \cite{SOC_paper}. We note this rotation does not break any symmetry, as the skyrmion state already has a preferred handedness.  

Near the upper critical field of the skyrmion phase, the magnetic system is near the threshold for conical spin alignment, where the Zeeman energy overcomes the $M^4$ energy minimization that occurs in the skyrmion state. The sublattice rotation preserves the skyrmion state, and hence the decrease in energy of the $M^4$ term, while still minimizing the Zeeman energy. Additionally, the long-range nature of the fluctuation minimizes distortion of the ground-state spiral structure.  The fluctuations are long-range and therefore conserve the local spin symmetry, indicating that the interaction that causes the rotation is smaller than the interactions that stabilize the skyrmion phase.

To gain insight into the slowly fluctuating nature of the skyrmion lattices, we have performed time-lapsed x-ray scattering measurements. After stabilizing the sample temperature to 57.5 K  $\pm$ 25 mK, we took images of the skyrmion peaks at 20 second intervals to track the thermally induced motion of the structure.  In Fig 5 (a - d) a series of snapshots every 15 minutes are shown. Interestingly, we found that both the helical as well as the skyrmion peak show rotational motion as a function of time. We note that this observation is quite different from what has been reported so far, where either an electric current or temperature gradient was identified as the reason for the motion \cite{Jonietz, Everschor, SekiPRB}. We observe the rotational motion at an equilibrium temperature and magnetic field, suggesting that the photocurrent due to x-rays could be the cause. 

\begin{figure}
\label{fig:fig5}
\includegraphics[width=3.25in]{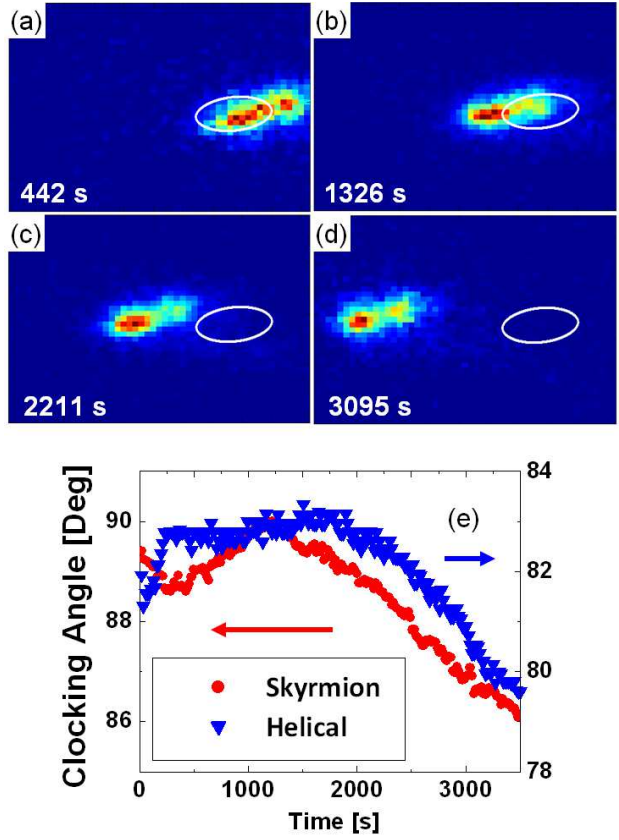}
\caption{(a - d) Camera images of skyrmion peak at fixed temperature and field.  The white oval indicates the position of the peak at frame 1. (e) Time scans of the clocking angle of the skyrmion and helical structures. }
\end{figure}

As shown in Fig 5(e), we have plotted the in-plane angle of one skyrmion and one helical peak as a function of time.  The skyrmion peak shows a counter-clockwise rotation (200 to 1200 s) before a reversing direction and rotating clockwise (1200 to 3500 s). The motions are not continuous but rather start and stop abruptly. For the skyrmions, we found that the sense of rotational motion can even change direction, which points to the intermittent nature of the dynamics. During the intermittent rotation of the skyrmion lattice, the relative angular separation between sublattice peaks does not change; both the peaks move in unison. This indicates the energy scale of the peak splitting is larger than the skyrmion lattice anisotropy energy. 

The observation of skyrmions with x-rays immediately opens up new class of science involving equilibrium and non-equilibrium behavior of the skyrmions. Particularly interesting will be chemical contrast specific measurements towards understanding the mechanisms to control and reconfigure the rotated skyrmion states to create new topological phases. Dynamical x-ray measurements will allow for study of the electric-field induced changes to the skyrmion lattice, relevant to spintronic applications, and time-domain measurements of the forced annihilation of skyrmions and corresponding creation of monopoles, e.g. through laser excitation.

\begin{acknowledgments}
The work at LBNL, including experiments at ALS, was supported by the Director, Office of Science, Office of Basic Energy Sciences, of the U.S. Department of Energy under Contract No. DEAC02-05CH11231. The work of J.L. and X.S. was partially supported by the U.S. Department of Energy, Office of Basic Energy Sciences, Division of Materials Science and Engineering under Grant No. DE- FG02-11ER46831.
\end{acknowledgments}

\chapter{}
\include{skyrmion_v6_supplemental}

\end{document}